\documentclass[aps,preprint,showpacs,showkeys]{revtex4}
\usepackage{amsmath}
\usepackage{amssymb}
\usepackage{graphicx}

\begin{document}

\title{ Broad class of nonlinear Langevin equations with  drift and diffusion coefficients separable in time and space: Generalized $n$-moment, ergodicity, Einstein relation and fluctuations of the system}
\author{Kwok Sau Fa}
\affiliation{Departamento de F\'{\i}sica, Universidade Estadual de Maring\'{a}, Av. Colombo 5790, 87020-900, \ Maring\'{a}-PR, Brazil, Tel.: 55 44 32614330}
\email{kwok@dfi.uem.br}

\author{Salete Pianegonda}
\affiliation{Department of Physics, Universidade Federal do  Paran\'{a},  Curitiba-PR, Brazil}


\begin{abstract}
A wide class of nonlinear Langevin equations with  drift and diffusion coefficients separable in time and space driven by the Gaussian white noise is analyzed in terms of a generalized n-moment. We show the system may present ergodic property, a key property in statistical mechanics, for space-time-dependent drift and diffusion coefficients. A generalized Einstein relation  is also  obtained. Besides, we show that the first two generalized moments and variance are useful to describe the drift and fluctuations of the system.
\end{abstract}

\maketitle
\newpage

\section{ Introduction}

Analyses in non-equilibrium statistical systems involve different quantities such as moments and their relations, ergodicity, correlation functions and Einstein relation \cite{risken,kubo,coffey,gitter,snook,moss,chers}. The probability density function (PDF) is a fundamental quantity from which we can  calculate any  quantity of interest. In fact, the PDF contains all the information of a system. Non-equilibrium statistical systems have been applied to  physics and   different   interdisciplinary areas \cite{coffey,snook,moss}, for instance, dynamics in graphene-based Josephson junctions \cite{spag3}, activation energies of oxygen ion diffusion in yttria stabilized zirconia by flicker noise spectroscopy \cite{spag4},  population genetics \cite{kimura}, spike train statistics for consonant and dissonant musical accords \cite{spag5}, a memristor stochastic model \cite{spag6}, relativistic Brownian motion \cite{dunkel} and quantum systems \cite{spag8}. 
 In particular, a key property  in statistical mechanics is the  ergodic property in which  the mean-squared displacement (MSD) is equivalent to  the  time-averaged mean-squared displacement (TAMSD); the practical interest  of the ergodic property involves the information obtained from time averages of a single or few trajectories which may represent  the ensemble averages. However, many statistical  systems exhibit  ergodicity breaking; there are only  a few systems that  present ergodic property.
 For example, the ergodicity breaking  is observed in a Langevin equation  with non-constant diffusion coefficients  \cite{chers} and colored noise \cite{yong}, however diffusion on random fractals presents ergodic property \cite{meroz,meroz2}. Fractional Brownian motion and fractional Langevin equation  driven by long-range correlated Gaussian noise are transiently non-ergodic \cite{jeon,jeonB} in which time-averaged quantities behave differently from their ensemble-averaged counterparts even for ergodic systems.  Ergodicity breaking has also been observed for continuous time random walks model \cite{montroll,montrollB} with diverging characteristic waiting time  \cite{bouch,bouchB,bouchC}, the blinking dynamics of quantum dots  \cite{steph}, protein motion in human cell walls \cite{weig}, lipid granule diffusion in yeast cells \cite{jeon2} and of insulin granules in MIN6 cells \cite{tabei}.
In fact, these recent investigations have shown ergodicity breaking in most of the systems analyzed. An other interesting quantity is the Einstein relation which connects the first moment in the presence of a constant force to the second moment without any external force, i.e, it connects the fluctuations of an ensemble of particles with their mobility under an applied small load force; it  can be obtained from the linear response theory \cite{barkai},  uncoupled continuous time random walk model \cite{barkai,kwokERc} and the Boltzmann transport equation \cite{mars}. The Einstein relation is also associated with the first fluctuation-dissipation theorem in which the mobility is described in terms of the equilibrium velocity correlation function (using the linear response theory) \cite{kubo,pott}. The Einstein relation has been verified by experiments in photocarrier drift and ambipolar diffusion in semiconductors \cite{gu} and in viscoelastic system \cite{amb}.

The aim of the work is to consider a wide class of Langevin equations with the drift and diffusion coefficients separable in time and space driven by the Gaussian white noise in the Stratonovich prescription. We investigate the ergodicity, Einstein relation and fluctuations of the system in terms of a generalized $n$-moment. We show that the system  may present the ergodic property for space-time-dependent drift and diffusion coefficients, and a generalized Einstein relation may also be obtained. Besides, we  show that the first two generalized moments and  generalized variance are useful to describe the  drift and fluctuations of the system. We also include two systems with time-dependent drift and diffusion coefficients into the discussions.

\section{Preliminary}

We consider the following nonlinear Langevin equation with the drift and diffusion coefficients separable in time and space driven by the  Gaussian white noise (in the Stratonovich approach), in one-dimensional space:

\begin{equation}
\frac{dx}{dt}= \left( \beta (t)+\gamma (t)G(x)\right) D(x)+b(t)  D(x) L(t)  ,  \label{eq1P}
\end{equation}%
where  $\beta (t)$, $\gamma (t)$ and $b(t)$ are functions of $t$,  $D(x)$ is a function of the variable $x$, $G(x)$ is described by
\begin{equation}
\frac{dG(x)}{dx}= \frac{1}{D(x)}  ,  \label{eq1aP}
\end{equation}%
and $L(t)$ is  the white noise force  with the following averages \cite{risken}:
\begin{equation}
\left\langle L (t)\right\rangle =0 \label{eqL1P}
\end{equation}%
and 
\begin{equation}
\left\langle L (t)L (\overline{t})\right\rangle =2\delta (t-\overline{t}) , \label{eqL2P}
\end{equation}%
 where $\delta (z)$ is the Dirac delta function. 
We also consider $b(t)$ and $D(x)$  non-negative functions. Eq. (\ref{eq1P}) has been used
 for various situations such as investigation of  turbulent two-particle diffusion in configuration space \cite{richar,richarb,richarc,richard,richare}, ergodic property \cite{chers,cher2,cher3,hou,cher4,leibo,wang1,wang2}, stochastic systems with resetting \cite{sandev,monteiro}, stochastic population dynamics \cite{calisto2,popul,kwok3} and the fixed samples of the mouse brain using magnetic resonance imaging \cite{magin,kwok4}. It can also describe interesting behaviors such as the asymptotic shape of the random-walk model \cite{kwok1} and logarithmic oscillations for the second moment \cite{kwok2}. 
A solution for the PDF and generalized $n$-moment of the above model have been obtained in Ref. \cite{kwok2020}. We present below a summary of the results.

The corresponding Fokker-Planck equation for the Langevin equation (\ref{eq1P}) in the Stratonovich approach is given by \cite{risken}

\begin{equation*}
\frac{\partial \rho(x,t)}{\partial t}=-\frac{\partial }{\partial x}\left[ \left( \beta (t)+\gamma(t) G(x)\right) D(x) \rho (x,t)\right]+ 
\end{equation*}%
\begin{equation}
b^2(t)\frac{\partial }{\partial x} \left[ D(x) \frac{\partial D(x) \rho (x,t)}{\partial x}\right] ,  \label{eq2P}
\end{equation}%
where  $\rho\left( x,t\right)$ is the PDF.

 For solving Eq.  (\ref{eq2P}) we consider the following initial condition \cite{kwok2020}:
\begin{equation}
 W(G(x),t=0)=\delta \left( G(x)-G_0\right)  , \label{eq2PB}
\end{equation}%
where $W(x,t)=D(x)\rho (x,t)$ and $G_0=G(t=0)$.
The  PDF has been obtained by using transformations of variables and the Fourier transforms \cite{kwok2020}, and it is  given by
\begin{equation}
\rho \left( x,t\right) =C\frac{\exp \left[ -\frac{\left( G(x)-M(t)\right)^2}{2\sigma (t)}\right] }{\sqrt{\sigma (t)}  D(x) } , \label{eq15P}
\end{equation}%
where  $C$ is a normalization constant and the coefficients $M\left( t\right)$ and $\sigma \left( t\right)$ are given by
\begin{equation}
M\left( t\right) =G_0 e^{\int_{0}^{t} \text{d}u \gamma (u)}+\int_{0}^{t} \text{d}z \beta (z) e^{\int_{z}^{t} \text{d}u \gamma (u)} \label{eq7aP}
\end{equation}%
and
\begin{equation}
\sigma \left( t\right) =2\int_{0}^{t} \text{d}z b^2(z) e^{2\int_{z}^{t} \text{d}u \gamma(u)} . \label{eq7bP}
\end{equation}%
 The coefficient $M(t)$ contains only the drift coefficients $\beta (t)$ and $\gamma (t)$. However, the coefficient $\sigma (t)$  contains  the diffusion coefficient $b(t)$ and the drift coefficient $\gamma (t)$ due to   $G(x)$ in Eq. (\ref{eq1P}). It can be seen that the coefficient $\sigma(t)$ given by Eq. (\ref{eq7bP})  is a positive quantity. 

The PDF  (\ref{eq15P}) can be normalized, however, the interval of the space will depend on the specific system; for example,  the variable $x$ in the population growth models is  interpreted as the number of population alive at time $t$, which is a non-negative quantity. In particular, the ordinary $n$-moment can  be obtained  for some specific forms of $D(x)$  \cite{kwok2019}.  But, we can consider the generalized $n$-moment given by $<G^n(x)>$ which recovers the ordinary $n$-moment for $D(x)=1$; in this case we can calculate the generalized $n$-moment for generic $D(x)$. We also consider  the whole space (-$\infty$, $\infty$),  $\left( G(-\infty)-M(t)\right) / \sqrt{ \sigma (t)} \rightarrow -\infty$ and $\left( G(\infty)-M(t) \right) / \sqrt{ \sigma (t)}\rightarrow \infty$ with the bounded functions for $M(t)$ and $\sigma (t)$, then the normalized PDF is given by
\begin{equation}
\rho \left( x,t\right) =\frac{\exp \left[ -\frac{\left( G(x)-M(t)\right)^2}{2 \sigma (t)}\right] }{ \sqrt{2 \pi \sigma (t)}  D(x) } . \label{eqcase61}
\end{equation}%
 From the above PDF we can obtain the generalized $n$-moment given by
\begin{equation*}
\left< G^n(x)\right> =\int_{-\infty}^{\infty} \text{d}x G^n(x)\rho \left( x,t\right)  =\sqrt{\frac{\left( 2 \sigma (t)\right)^n}{\pi}}\int_{-\infty}^{\infty} \text{d}u \left( u+\frac{M(t)}{\sqrt{2 \sigma (t)}}\right)^ne^{-u^2}
\end{equation*}%
\begin{equation}
=\frac{M^n(t)}{2\sqrt{\pi}}\sum_{k=0}^{n}\left( \begin{array}{c} n \\ k \end{array}\right) \left( \frac{\sqrt{2 \sigma (t)}}{M(t)}\right)^k\left( 1+(-1)^k\right) \Gamma \left( \frac{1+k}{2}\right) , \label{eqcase62}
\end{equation}%
where $\Gamma(z)$ is the gamma function.

For instance, the first two  generalized moments are given by
\begin{equation}
\left< G(x)\right> =M(t) , \label{eqcase63}
\end{equation}%
and
\begin{equation}
\left< G^2(x)\right> =\sigma (t)+M^2(t) . \label{eqcase64}
\end{equation}%
The generalized variance can be obtained from the first two generalized moments which yields
\begin{equation}
\left< \left( G(x)-\left< G(x)\right>\right)^2\right> =\sigma (t) . \label{eqcase64}
\end{equation}%
Note that the PDF and the generalized $n$-moment can be written in terms of the first generalized moment and generalized variance.

 The choice of the observable $G(x)$ in Eq.  (\ref{eqcase62}) has various advantages. First we can reduce Eq. (\ref{eq1P}) to a stochastic system with time-dependent linear force plus a time-dependent load force as follows:
\begin{equation}
\frac{d\bar{x}}{dt}=  \beta (t)+\gamma (t)\bar{x} +b(t)   L(t)  ,  \label{eqcase65}
\end{equation}%
where $\bar{x}(x)=G(x)$. Moreover, $G(x)$ can be used to obtain the ergodic property of some time-dependent systems described by Eq. (\ref{eq1P}) for generic $D(x)$, which is in contrast to the observable $x$. The function $G(x)$ can also be used to obtain a generalized Einstein relation. These quantities will be discussed in the next sections.

\section{Ergodicity}

Recently the ergodic property has been investigated in many different systems,  and they presented the ergodicity breaking in most of the cases. It is also difficult to find the ergodic property of the system (\ref{eq1P}) for generic drift and diffusion coefficients. However,  the ergodicity of a wide class of systems, with generic diffusion coefficient, may be attained by using the second generalized moment and generalized time average \cite{kwokCSF2022}.  For the system (\ref{eq1P}) we can obtain expressions for the generalized correlation function, generalized mean squared displacement (GMSD) and generalized ensemble time average  of a single-particle tracking (GETAMSD). As example, we will show that the system (\ref{eq1P}), with $\beta (t)= \beta$, $\gamma (t)=-\gamma$ and $b(t)=\sqrt{b}$, can  present ergodic property by using the GMSD and GETAMSD; this system generalizes the result obtained in Ref. \cite{kwokCSF2022}.

The PDF (\ref{eqcase61}) is non-Gaussian due to the non-constant coefficient $D(x)$ and it  can describe anomalous diffusion processes \cite{kwok2}; but it is Gaussian for $D(x)=1$. 
For non-constant $D(x)$ the system can describe  different  behaviors from those described by $D(x)=1$. For instance, the system (\ref{eq1P}) for $\beta (t) =\gamma (t)=0$, $b(t)=\sqrt{D}$ and $D(x)= \vert x\vert^{\nu /2}$ with $\nu < 2$ may present bimodal distribution, and its second moment   is given by
\begin{equation}
\left< x^2(t) \right>=\frac{\left( D\left( 2-\nu \right)^2\right)^{\frac{2}{2-\nu}}\Gamma\left( 1+\frac{2+\nu}{2(2-\nu )}\right) t^{\frac{2}{2-\nu}}}{\sqrt{\pi}}  , \label{eqSol1d}
\end{equation}
where $x_0=x(t=0)=0$.
Eq. (\ref{eqSol1d})  describes subdiffusion ($\nu <0$), normal diffusion ($\nu =0$) and superdiffusion ($\nu >0$) processes, with the  ballistic process given by $\nu =1$.

 The ergodicity of a system  is established by  the time average of a single-particle tracking and the MSD which corresponds to an ensemble of particles.  The time average of a single-particle tracking (TAMSD) is defined by
\begin{equation}
\overline{\delta^2( t ) }=\frac{1}{T-t}\int_0^{T-t}\left[ x\left( \tau +t \right)-x(\tau ) \right]^2 d\tau ,  \label{eqTAM}
\end{equation}
where $t$ is the lag time and $T$ is the measurement time; whereas the MSD is defined by
\begin{equation}
\left< x^2(t) \right>=\int_{-\infty}^{\infty}x^2 \rho \left( x,t\right) dx  . \label{eqTAMb}
\end{equation}
The ergodic property of a system is established when the  TAMSD  converges to the MSD  in the long measurement time limit, i.e,  $\lim_{T\rightarrow \infty}\overline{\delta^2( t ) }=\left< x^2(t) \right>$ \cite{maria}.  Usually, we employ the ensemble of $\overline{\delta^2( t ) }$ which is  smoother than $\overline{\delta^2( t ) }$.

The ergodicity of the system (\ref{eq1P}) has been investigated in Ref. \cite{chers}  for $\beta (t) =\gamma (t)=0$, $b(t)=\sqrt{D}$, $x_0=0$ and $D(x)= \vert x\vert^{\nu /2}$ with $\nu < 2$, and the result is given by 
\begin{equation}
\left< \overline{\delta^2( t ) }\right> =\left( \frac{t}{T}\right)^{1-p}\left< x^2(t )\right> ,  \ \  t \ll T, \label{eqTAM2}
\end{equation}
where $p=2/(2-\nu)$ and $\left< \overline{\delta^2( t ) }\right>$ is the ensemble TAMSD (ETAMSD). The ETAMSD  is described by  a sum of individual trajectories given by $\left< \overline{\delta^2( t ) }\right> = N^{-1}\sum_{i=1}^N\overline{\delta_i^2( t ) }$, where $N$ is the number of  individual trajectories.  Eq. (\ref{eqTAM2}) shows ergodicity breaking due to the fact that the ETAMSD and the MSD are not equivalent except for the Wiener process ($p=1$); thus the  Langevin equation  with the space-dependent diffusion coefficient $D(x)= \vert x\vert^{\nu /2}$ does not present  the ergodic property, in the Stratonovich prescription. The   ergodicity breaking is also observed for other functions  \cite{chers}. In general, we will find the ergodicity breaking in the system (\ref{eq1P}) for non-constant $D(x)$. 

However, the ergodic property of the system (\ref{eq1P}) can be obtained, at least for some class of drift and diffusion coefficients,  by using generalizations of the MSD and TAMSD, i.e, we replace the observable  $x(t)$ by $G(x)$.  In the case of TAMSD, we replace  the ordinary squared displacement $\left[ x\left( \tau +t\right)-x(\tau ) \right]^2$ by a generalized squared displacement given by $\left[ G\left( \tau +t \right)-G(\tau ) \right]^2$. In this case, the generalized TAMSD (GTAMSD) is given by
\begin{equation}
\overline{\delta_G^2( t ) }=\frac{1}{T-t}\int_0^{T-t}\left[ G\left( \tau +t \right)-G(\tau ) \right]^2 d\tau , \label{eqTAM3}
\end{equation}
where the lower index $G$ on the left-hand side indicates that the  ordinary squared displacement is replaced by the generalized squared displacement.  The GTAMSD may or may not depend on the initial value $G_0$, then the generalized MSD (GMSD) would be chosen according to  specific systems in order to attain the ergodic property, i.e., $\left< \left( G(x)-A(t)\right)^2 \right>$, where $A(t)$ would be chosen according to specific systems; the function $A(t)$ generalizes the initial value $x_0$ in the ordinary MSD ($\left< \left( x-x_0\right)^2 \right>$). For  systems described below we choose the  GMSD as follows: 
\begin{equation*}
\left< \left( G(x)-G_0 e^{\int_{0}^{t} \text{d}u \gamma (u)}\right)^2 \right> =\left<  G^2(x) \right> -2G_0 e^{\int_{0}^{t} \text{d}u \gamma (u)}\left<  G(x) \right> +G_0^2 e^{2\int_{0}^{t} \text{d}u \gamma (u)}= 
\end{equation*}%
\begin{equation}
\sigma \left( t\right)+M^2(t)-2G_0 e^{\int_{0}^{t} \text{d}u \gamma (u)} M(t)+G_0^2 e^{2\int_{0}^{t} \text{d}u \gamma (u)}=\sigma \left( t\right)+\bar{M}^2(t) , \label{eqTAM4}
\end{equation}%
where $A(t)=G_0 e^{\int_{0}^{t} \text{d}u \gamma (u)}$ and $\bar{M}(t)$ is given by
\begin{equation}
\bar{M}(t)=\int_{0}^{t} \text{d}z \beta (z) e^{\int_{z}^{t} \text{d}u \gamma (u)} . \label{eqTAM4b}
\end{equation}%
The difference between  $\bar{M}(t)$ and  $M(t)$ is that $\bar{M}(t)$ does not contain the initial value $G_0$.

 It should be noted that the GTAMSD recovers the  ordinary TAMSD for $D(x)=1$. In particular, the ergodic property of the system (\ref{eq1P}) has been verified by using the  GTAMSD and GMSD for  $\beta (t) =\gamma (t)=0$, $b(t)=\sqrt{D}$ and $D(x)= \vert x\vert^{\nu /2}$ with $\nu < 2$ \cite{kwokCSF2022}.

In order to calculate the GETAMSD  for generic $D(x)$,  from Eq.   (\ref{eqTAM3}), we need the generalized correlation function $\left< G(\tau +t)G(\tau ))\right>$ which  is defined by \cite{risken}
 \begin{equation}
\left< G(t_2)G(t_1 ))\right> =\int_{-\infty}^{\infty}dx_2\int_{-\infty}^{\infty}dx_1 G\left( x_2\right) G\left( x_1\right) P(x_2,t_2\vert x_1,t_1) \rho(x_1,t_1), \ \  t_2>t_1 , \label{eqTAM5}
\end{equation}
where $P(x_2,t_2\vert x_1,t_1)$ is the transition probability \cite{risken} given by
 \begin{equation}
 P(x_2,t_2\vert x_1,t_1) =\frac{e^{ -\frac{ \left[ G(x_2)-G(x_1) e^{\int_{t_1}^{t_2}du \gamma (u)}-\bar{M}(t_2,t_1) \right]^2}{2\sigma \left( t_2,t_1\right)}}}{\sqrt{ 2\pi \sigma \left( t_2,t_1\right) } D(x_2)  }  , \ \  t_2>t_1 , \label{eqTP}
\end{equation} 
 \begin{equation}
\bar{M}(t_2,t_1)= \int_{t_1}^{t_2}dz \beta \left( z\right) e^{\int_{z}^{t_2}du \gamma \left( u\right)} \label{eqTP2}
\end{equation}
and
\begin{equation}
\sigma (t_2,t_1)=2 \int_{t_1}^{t_2}dz b^2 \left( z\right) e^{2\int_{z}^{t_2}du \gamma \left( u\right)} .\label{eqTP2}
\end{equation}
It should be noted that   $\bar{M}(t)=\bar{M}(t,0)$ and $\sigma(t)=\sigma(t,0)$.
Substituting Eqs.  (\ref{eqcase61}) and (\ref{eqTP}) into Eq.  (\ref{eqTAM5}) yields
 \begin{equation*}
\left< G(\tau +t)G(\tau ))\right> =e^{\int_{\tau}^{t+\tau}du \gamma \left( u\right)}\left[ \sigma \left( \tau ,0 \right)+\left( G_0 e^{\int_{0}^{\tau}du \gamma \left( u\right)}+\bar{M}(\tau,0)\right)^2\right] +
\end{equation*}
 \begin{equation}
\bar{M}(\tau +t,\tau) \left[ G_0 e^{\int_{0}^{\tau}du \gamma \left( u\right)}+\bar{M}(\tau,0) \right] . \label{eqTAM5b}
\end{equation}
From Eq. (\ref{eqTAM3}) we obtain the following expression for the GETAMSD:
\begin{equation*}
\left< \overline{\delta_G^2( t ) }\right> =\frac{1}{T-t}\int_0^{T-t}\left[ \left< G^2\left( \tau +t \right) \right>-2\left< G\left( \tau +t\right) G(\tau ) \right> +\left< G^2\left( \tau \right) \right>\right] d\tau =
\end{equation*}
\begin{equation*}
\frac{1}{T-t}\int_0^{T-t}\left\{ \sigma \left( \tau +t  ,0 \right)+\left( G_0 e^{\int_{0}^{\tau +t}du \gamma \left( u\right)}+\bar{M}(\tau +t,0)\right)^2 - \right.
\end{equation*}
\begin{equation*}
\left. 2\bar{M}(\tau +t,\tau) \left[ G_0 e^{\int_{0}^{\tau}du \gamma \left( u\right)}+\bar{M}(\tau,0) \right] +\right.
\end{equation*}
\begin{equation}
\left. \left( 1-2e^{\int_{\tau}^{\tau +t}du \gamma \left( u\right)}\right) \left[ \sigma \left( \tau ,0 \right)+\left( G_0 e^{\int_{0}^{\tau}du \gamma \left( u\right)}+\bar{M}(\tau,0)\right)^2 \right] \right\} d\tau . \label{eqTAM5c}
\end{equation}
Eq. (\ref{eqTAM5c}) is the GETAMSD for generic $\beta(t)$, $\gamma(t)$ and $b(t)$,  which may depend on the initial value $G_0$.

Next we take two examples of the generalized correlation function, GMSD and GETAMSD.

{\it Case 1.} $\beta(t) =\gamma (t) =0$ and $b(t)=\sqrt{D}$.

This case has been analyzed in Ref. \cite{kwokCSF2022}. From Eqs. (\ref{eqTAM4}), (\ref{eqTAM5b}) and (\ref{eqTAM5c}) we obtain the following  generalized correlation function, GMSD and GETAMSD:
 \begin{equation}
\left< G(\tau +t)G(\tau ))\right> =2D\tau +G_0^2 , \label{eqTAM5d}
\end{equation}
 \begin{equation}
\left< \left( G(x)-G_0\right)^2\right> =2Dt \label{eqTAM5e}
\end{equation}
and
\begin{equation}
\left< \overline{\delta_G^2( t ) }\right> =2 D t  . \label{eqTAM5f}
\end{equation}
  Eq.  (\ref{eqTAM5f}) shows that the GETAMSD does not depend on the initial value $G_0$. We can see that  the GMSD and GETAMSD given by Eqs.  (\ref{eqTAM5e}) and  (\ref{eqTAM5f}) scale linearly  and they also recover the results of the Wiener process for $D(x)=1$.  Comparing Eq.  (\ref{eqTAM5e}) with Eq.  (\ref{eqTAM5f}) yields
\begin{equation}
\left< \overline{\delta_G^2( t ) }\right> =2Dt    =\left< \left( G(x )-G_0\right)^2\right>  . \label{eqTAM6}
\end{equation}
Eq.  (\ref{eqTAM6}) shows that the ergodicity is attained for the system (\ref{eq1P}) by using modified MSD and TAMSD. 

{\it Case 2.} $\beta(t) =\beta $, $\gamma (t) =-\gamma $ and $b(t)=\sqrt{b}$.

In this case  the  generalized correlation function, GMSD and GETAMSD are described by
 \begin{equation*}
\left< G(\tau +t)G(\tau ))\right> =e^{-\gamma t}\left[ \frac{b}{\gamma}  \left( 1-e^{-2\gamma \tau}\right) +\left( G_0e^{-\gamma \tau}+\frac{\beta}{\gamma} \left( 1-e^{-\gamma \tau }\right) \right)^2 \right] +
\end{equation*}
\begin{equation}
\frac{\beta}{\gamma}\left( 1-e^{-\gamma t}\right)\left( G_0e^{-\gamma \tau}+\frac{\beta}{\gamma} \left( 1-e^{-\gamma \tau }\right) \right)
 , \label{eqTAM5d}
\end{equation}
 \begin{equation}
\left<  \left( G(x)-G_0e^{-\gamma t}\right)^2\right> =\frac{b}{\gamma} \left( 1-e^{-2\gamma t}\right) +
\frac{\beta^2}{\gamma^2} \left( 1-e^{-\gamma t}\right)^2
 \label{eqTAM7}
\end{equation}
and
\begin{equation}
\left< \overline{\delta_G^2( t ) }\right> =\frac{2b}{\gamma}
 \left( 1-e^{-\gamma t}\right) +
\frac{1}{T-t}\left[ \frac{b}{2\gamma^2}- \frac{\left( G_0-\frac{\beta}{\gamma}\right)^2}{2\gamma}\right]  \left( 1-e^{-\gamma t}\right)^2 \left( e^{-2\gamma (T-t)}-1\right) .\label{eqTAM8}
\end{equation}
 The expressions for the GMSD and GETAMSD are very different and, consequently, the system does not present the ergodic property for  generic parameter values. Moreover, the GETAMSD given by Eq.  (\ref{eqTAM8})  depends explicitly on the initial value $G_0$ and the measurement time T, except the first term on the right-hand side. The ergodic property can be obtained  from Eqs.  (\ref{eqTAM7}) and  (\ref{eqTAM8}) by taking  $\beta =0$ and $\gamma \rightarrow 0$. For $D(x)=1$  the system recovers the Ornstein-Uhlenbeck process with the presence of a load force. The results (\ref{eqTAM7}) and  (\ref{eqTAM8}) for $D(x)=1$, $\beta =0$ and $x_0=0$ are in agreement with those presented in Ref. \cite{jeon3}. It should be noted that the ergodic property of the Wiener process is obtained  from Eqs.  (\ref{eqTAM7}) and  (\ref{eqTAM8}) by taking $D(x)=1$, $\beta =0$ and $\gamma \rightarrow 0$.

\begin{figure}
\includegraphics[width=15cm, height=10cm]{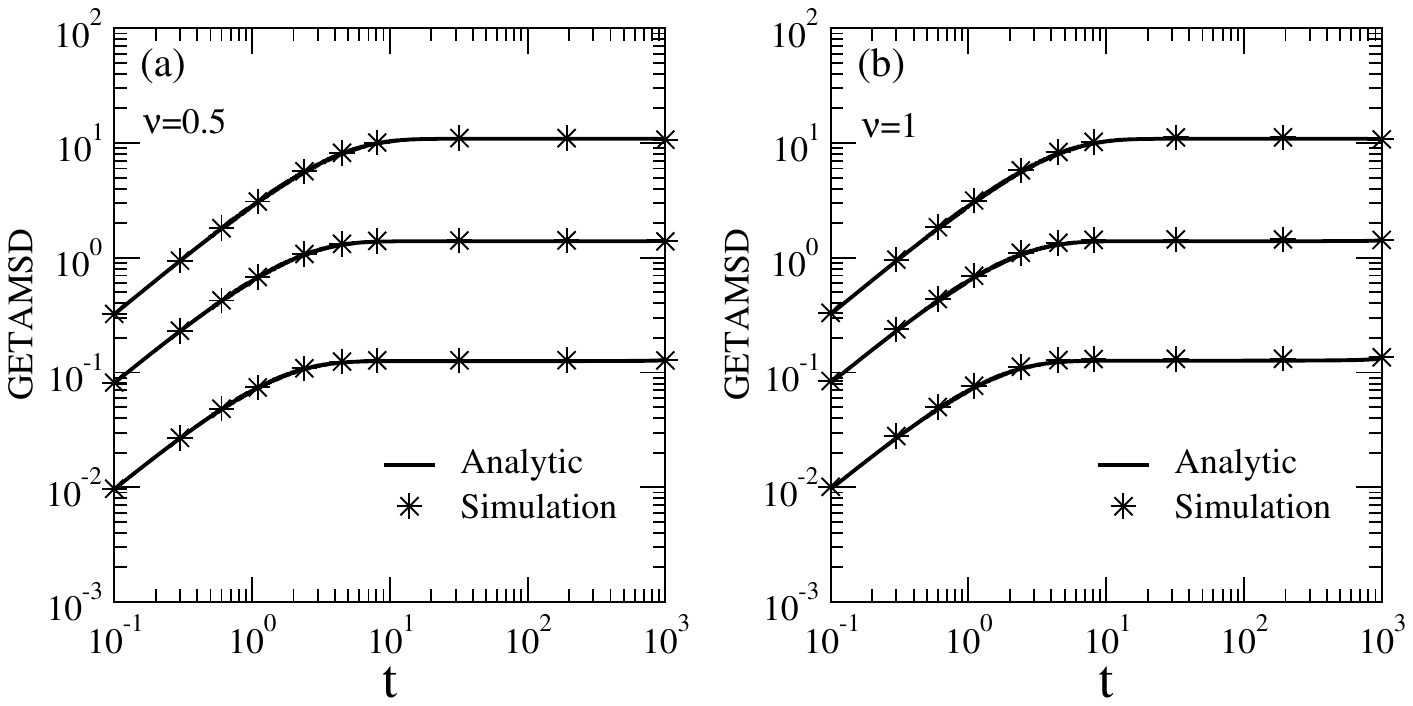}
\caption{\label{fig1} Log-log plots of the simulated and analytical results for the  GETAMSD with  $D(x)= \vert x\vert^{\nu /2}$. The analytical result is described by Eq. (\ref{eqTAM8}). We take the initial value $x_0=1$ and $T=1200$. The parameter values  ($\beta$, $\gamma $ and $b$) are the same for both figures ((a) and (b)); from top to bottom the parameter values are: $\beta=0.7$, $\gamma =0.3$, $b=\beta^2/\gamma=1.6333$; $\beta=-0.5$, $\gamma =0.6$, $b=\beta^2/\gamma=.4167$; $\beta=0.2$, $\gamma =0.8$, $b=\beta^2/\gamma=0.05$. In a) the parameter $\nu$ is given by $\nu =1/2$, and $\nu =1$ for  b).   }
\end{figure}

\begin{figure}
\includegraphics[width=15cm, height=10cm]{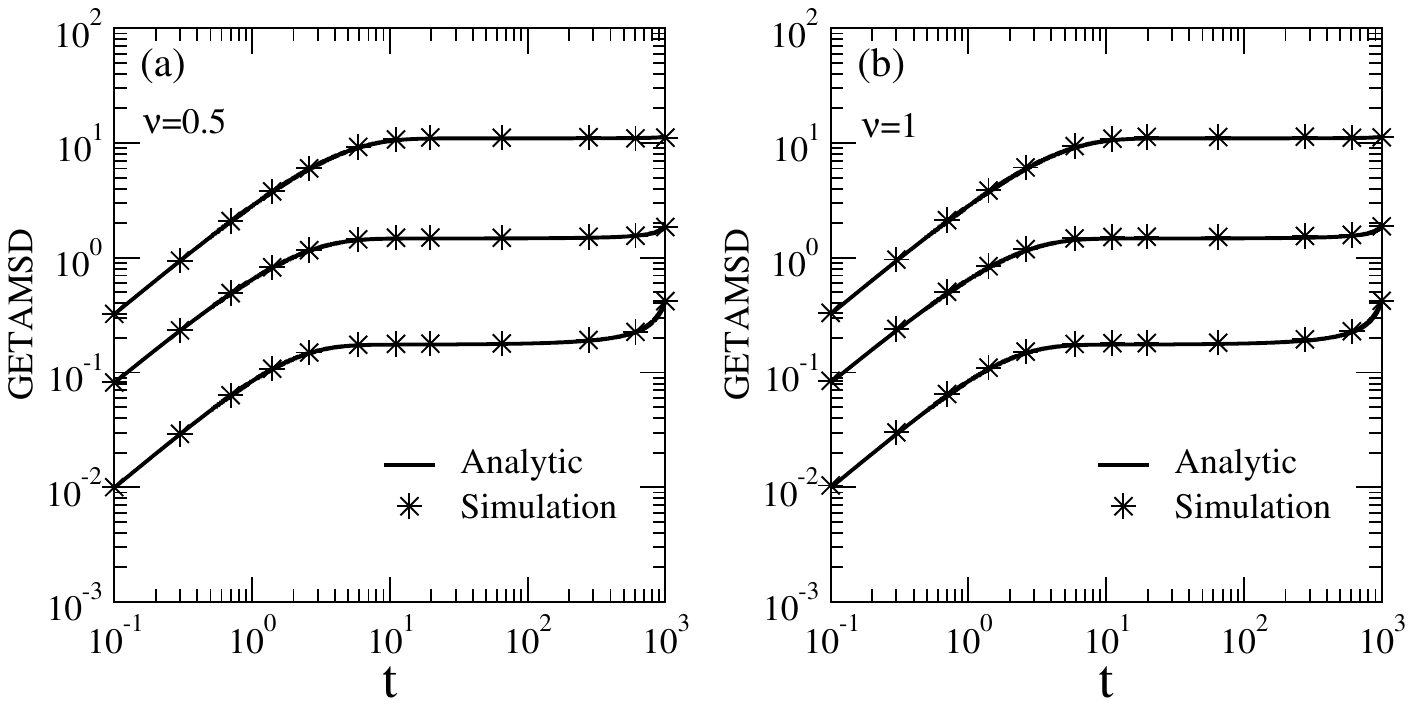}
\caption{\label{fig2} Log-log plots of the simulated and analytical results for the  GETAMSD with  $D(x)= \vert x\vert^{\nu /2}$. The analytical result is described by Eq. (\ref{eqTAM8}). We take the initial value $G_0=10$ and $T=1200$. The parameter values  ($\beta$, $\gamma $ and $b$) are the same for both figures ((a) and (b)); from top to bottom the parameter values are: $\beta=0.7$, $\gamma =0.3$, $b=\beta^2/\gamma=1.6333$; $\beta=-0.5$, $\gamma =0.6$, $b=\beta^2/\gamma=.4167$; $\beta=0.2$, $\gamma =0.8$, $b=\beta^2/\gamma=0.05$. In a) the parameter $\nu$ is given by $\nu =1/2$, and $\nu =1$ for  b).   }
\end{figure}

 It can be seen that, from Eqs.  (\ref{eqTAM7}) and  (\ref{eqTAM8}), the GMSD and GETAMSD present  plateaus due to the term $G(x)$ in Eq.  (\ref{eq1P}). 
It is interesting to note that $\beta$ does not generate a new term in Eq.  (\ref{eqTAM8}), but it is associated with the constants $b$ and $G_0$ which is in contrast to the result of the GMSD.
Moreover, from Eqs.  (\ref{eqTAM7}) and  (\ref{eqTAM8}), the system presents ergodicity breaking for $\beta =0$ and $\gamma \ne 0$. Figs. (\ref{fig1}) and (\ref{fig2}) show the simulated and analytical results for the GETAMSD with  different parameter values and $x_0\ne 0$. The analytical and theoretical results are in good agreement, and they do not depend on the $D(x)$. For comparison we have taken $x_0=1$ for Fig. (\ref{fig1}) and $G_0=10$ for Fig. (\ref{fig2}). They show  the behaviors of the GETAMSD which depend on the initial values. The curve in the bottom of  Fig. (\ref{fig1}) has the behavior changed by the initial value which is shown in  Fig. (\ref{fig2}).

For $t\ll 1$ and $T\gg t$ Eq.  (\ref{eqTAM8}) is given by
\begin{equation}
\left< \overline{\delta_G^2( t ) }\right> \approx \left< \left( G( x )-G_0 e^{- \gamma t}\right)^2 \right>+\frac{1}{\gamma}\left( b-\frac{\beta^2}{\gamma}\right) \left( 1-e^{-\gamma t}\right)^2 .  \label{eqAppr0}
\end{equation}
From Eq. (\ref{eqAppr0}) we can obtain the ergodic property  for  $t\ll 1$ and $T\gg t$ and $b=\beta^2/\gamma$. 

For $t\approx T$  Eq.  (\ref{eqTAM8}) is given   by
\begin{equation*}
\left< \overline{\delta_G^2(t ) }\right> \approx \frac{2b}{\gamma} \left( 1-e^{-\gamma t}\right) +\left[ \left( G_0-\frac{\beta}{\gamma}\right)^2-\frac{b}{\gamma}\right] \times 
\end{equation*}
 \begin{equation}
 \left(  1-e^{-\gamma t} \right)^2  \left[ 1-\gamma (T-t)\right]  .  \label{eqAppr1}
\end{equation}
 Eq. (\ref{eqAppr1}) can be written in terms of  $\left< (G( x )-G_0 e^{- \gamma t})^2 \right>$  as follows:
\begin{equation*}
\left< \overline{\delta_G^2( t ) }\right> \approx \left< \left( G(x)-G_0 e^{- \gamma t}\right)^2 \right>+\frac{1}{2\gamma}\left( G_0^2-\frac{2G_0\beta}{\gamma}\right)\left( 1-e^{-\gamma t}\right)^2 -
\end{equation*}
\begin{equation}
\gamma \left[ \left( G_0-\frac{\beta}{\gamma}\right)^2-\frac{b}{\gamma}\right] (T-t)\left( 1-e^{-\gamma t}\right)^2 .  \label{eqAppr2}
\end{equation}
Eq. (\ref{eqAppr2}) shows that the ergodic property can  be obtained from $G_0\ne 0$ with additional conditions $G_0=2\beta /\gamma$ and $b=\beta^2/\gamma$.
 Moreover,  the GETAMSD converges to  the GMSD for $b=\beta^2/\gamma $ and $G_0=0$, i.e, 
\begin{equation}
\left< G^2( x ) \right>=\left< \overline{\delta_G^2(t ) }\right> .  \label{eqErg}
\end{equation}
 For $D(x)=1$ the system describes the ergodicity of the Ornstein-Uhlenbeck process  restored by  a load force.

\begin{figure}
\includegraphics[width=15cm, height=10cm]{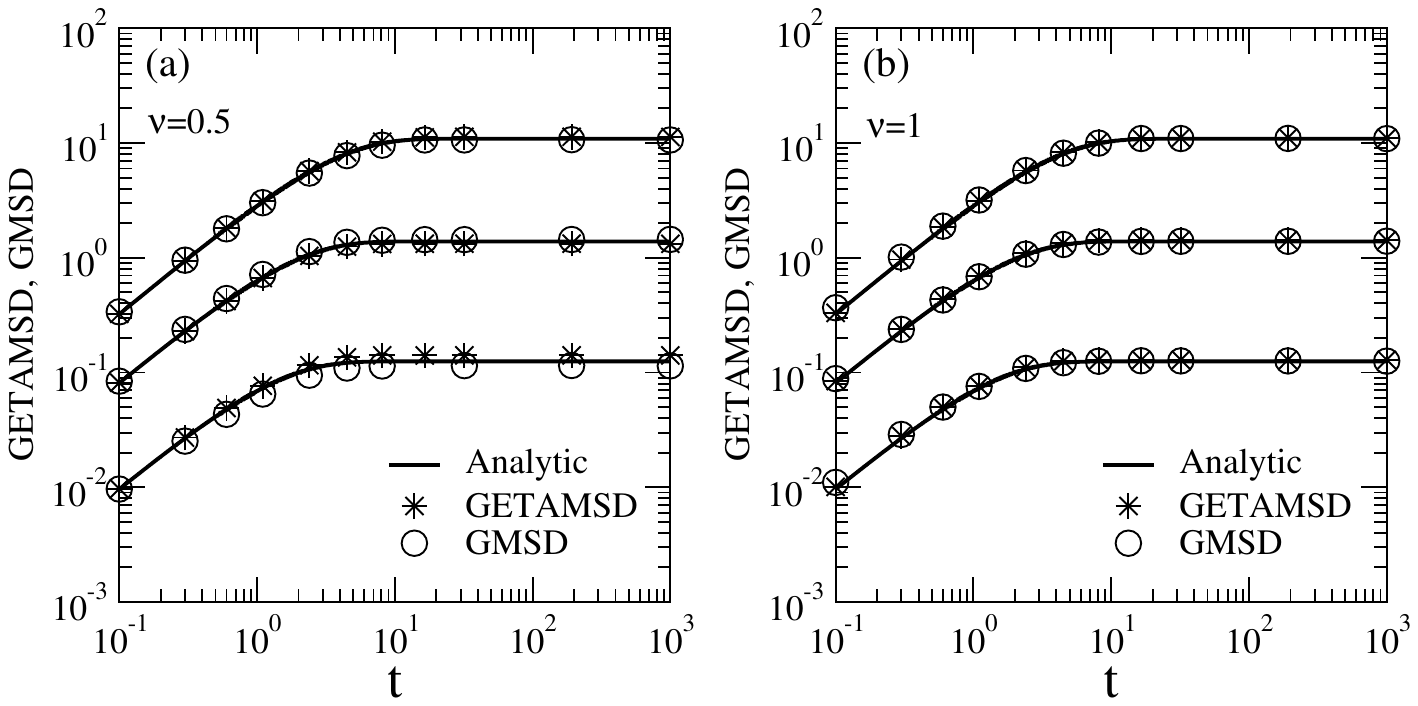}
\caption{\label{fig3} Log-log plots of the simulated and analytical results for the GMSD  and GETAMSD with  $D(x)= \vert x\vert^{\nu /2}$. The analytical result is described by Eq. (\ref{eqTAM9}). We take the initial value $x_0=0$ and $T=1200$. The parameter values  ($\beta$, $\gamma $ and $b$) are the same for both figures ((a) and (b)); from top to bottom the parameter values are: $\beta=0.7$, $\gamma =0.3$, $b=\beta^2/\gamma=1.6333$; $\beta=-0.5$, $\gamma =0.6$, $b=\beta^2/\gamma=.4167$; $\beta=0.2$, $\gamma =0.8$, $b=\beta^2/\gamma=0.05$. In a) the parameter $\nu$ is given by $\nu =1/2$, and $\nu =1$ for  b).   }
\end{figure}

The above behaviors for $D(x)=1$ are the same for generic $D(x)$ when we use the generalized moments.
In fact, the ergodic property of the system can be attained by taking $G_0=0$ and $\beta^2/\gamma =b$ or $T-t\gg 1$ and $\beta^2/\gamma =b$, and we have
\begin{equation}
\left< \overline{\delta_G^2( t ) }\right> =\frac{2b}{\gamma} \left( 1-e^{-\gamma t}\right)   =\left<  \left( G(x)+G_0e^{-\gamma t} \right)^2\right>  . \label{eqTAM9}
\end{equation}
It should be noted that the above ergodic property is valid for $G_0\ne 0$, with the conditions $T-t\gg 1$ and $\beta^2/\gamma =b$.
Eq.  (\ref{eqTAM9}) shows that the ergodic property is attained for the system (\ref{eq1P}) with generic $D(x)$  by using generalized MSD and TAMSD.  The expression (\ref{eqTAM9}) presents plateau, given by $2b/\gamma$, for $\gamma t \gg 1$, due to   $G(x)$ in Eq. (\ref{eq1P}). The  plateau disappears  for $\gamma \rightarrow 0$ and the expression (\ref{eqTAM9}) reduces to
\begin{equation}
\left< \overline{\delta_G^2( t ) }\right> =2bt   =\left< \left( G(x)-G_0\right)^2\right>  , \label{eqTAM10}
\end{equation}
which has the same result of the first case.

 Fig. \ref{fig3} shows  the behaviors of the GMSD and GETAMSD  for $D(x)$ described by a power-law; they present ergodic property with $b=\beta^2/\gamma$. Figs. \ref{fig3}a and \ref{fig3}b show behaviors  for the same parameter values ($\beta$, $\gamma$ and $b$), but  with different values of $\nu$. We can see the behaviors are independent of $D(x)$.  Moreover, the simulations are in very good agreement with the analytical results.

\section{ Generalized variance, fluctuations and generalized Einstein relation }

Fluctuations  represent an important part of a stochastic system. However, in many cases the fluctuations are not easily identified and calculated. For systems driven by multiplicative noise the fluctuations can not be separated from the drift using the first two moments due to the connection between the dependent variable and the noise.  For example, we consider $D(x)=x$  from which the space is limited to $0< x< \infty $ due to the normalization of the PDF;  this  system presents interesting results in which the  moments can be related among them. The coefficient $D(x)=x$ may also be useful to describe population growth models (see, for instance, \cite{roman}). In this case, the $n$-moment can be calculated exactly. From the PDF (\ref{eq15P}) we obtain 
\begin{equation}
 \left< x^n\right> =e^{nM(t)+\frac{n^2\sigma (t)}{2}} . \label{eq80}
\end{equation}
We can see that  the  time-dependent drift and diffusion coefficients in the  $n$-moment can not be separated as a sum due to the exponential function. Therefore, we can not obtain a quantity related to the fluctuations of the system by using the ordinary $n$-moment and variance. In order to separate $\sigma (t)$ from $M(t)$ we have to consider the following expressions:
\begin{equation}
 \frac{\left< x^{2n}\right> }{\left< x^n\right>^2}=e^{n^2\sigma (t)}  \label{eq80b}
\end{equation}
and
\begin{equation}
 \frac{\left< x^n\right>^4 }{\left< x^{2n}\right>}=e^{2nM (t)} . \label{eq80ba}
\end{equation}
We can see that Eq. (\ref{eq80b}) contains $n$ relations which depend only on the coefficient $\sigma (t)$, whereas Eq. (\ref{eq80ba}) contains $n$ relations which depend only on the  time-dependent drift coefficients.
For  $n=1$ Eqs.  (\ref{eq80b}) and (\ref{eq80ba}) reduce to
\begin{equation}
 \frac{\left< x^2\right> }{\left< x\right>^2}=e^{\sigma (t)}  \label{eq80bb}
\end{equation}
and\begin{equation}
 \frac{\left< x\right>^4 }{\left< x^2\right>}=e^{2M (t)} . \label{eq80bc}
\end{equation}
Eqs.  (\ref{eq80bb}) and (\ref{eq80bc}) are relations which contain only the first two moments. They can be written in terms of the relations (\ref{eq80b}) and (\ref{eq80ba}) as follows:
\begin{equation}
  \frac{\left< x^{2n}\right> }{\left< x^n\right>^2} =\left(  \frac{\left< x^2\right> }{\left< x\right>^2}\right)^{n^2} \label{eq80bd}
\end{equation}
and
\begin{equation}
 \frac{\left< x^n\right>^4 }{\left< x^{2n}\right>}= \left( \frac{\left< x\right>^4 }{\left< x^{2}\right>}\right)^{n} . \label{eq80be}
\end{equation}
Besides, we can obtain the following relation by using Eqs. (\ref{eq80bd}) and (\ref{eq80be}):
\begin{equation}
  \left< x^{n}\right>^2 = \left< x^2\right>^{n(n-1)} \left< x\right>^{2n(2-n)} . \label{eq80bda}
\end{equation}
Eqs. (\ref{eq80bd}),  (\ref{eq80be}) and (\ref{eq80bda}) show that the quantities $\left< x^{2n}\right> /\left< x^n\right>^2$,  $\left< x^n\right>^4 /\left< x^{2n}\right>$ and $\left< x^{n}\right>^2$  can be factorized in terms of the first two moments. 

\begin{figure}
\includegraphics[width=15cm, height=10cm]{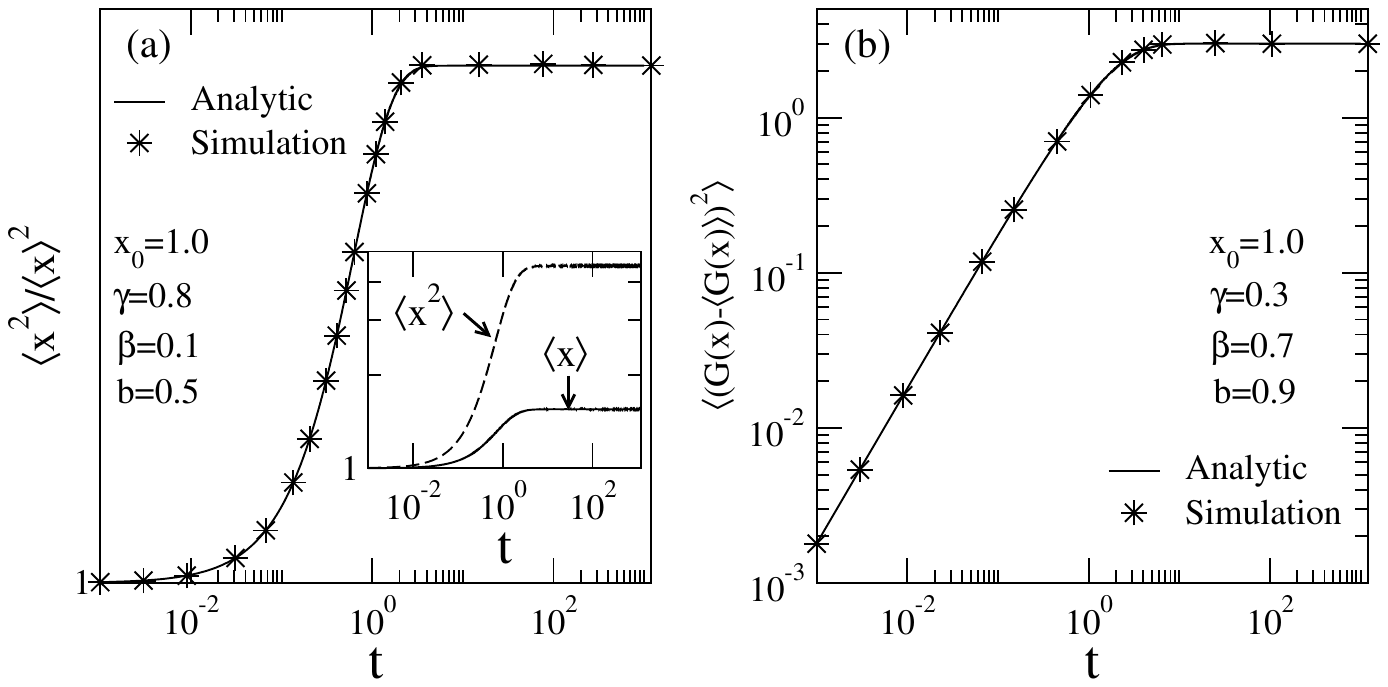}
\caption{\label{fig4}  Log-log plots of the simulated and analytical results for the relation  (\ref{eq80bb}) and the generalized variance described by Eq.  (\ref{eq83}). The time-dependent coefficients are given by $\beta (t) =\beta$, $\gamma (t)=-\gamma$ and $b(t)=\sqrt{b}$.  }
\end{figure}

For $\gamma (t)=0$ Eq.  (\ref{eq80b}) reduces to
\begin{equation}
 \frac{\left< x^{2n}\right> }{\left< x^n\right>^2}=e^{2n^2\int_{0}^{t} \text{d}z b^2(z)} , \label{eq80c}
\end{equation}
which depends only on the time-dependent diffusion coefficient $b(t)$.
 However, we can use the first generalized moment and generalized variance to quantify the drift and  fluctuations of the system,
\begin{equation}
 \left< G(x)\right> =M (t)  \label{eq83}
\end{equation}
and
\begin{equation}
 \left< \left( G(x)-\left< G(x)\right>\right)^2\right> =\sigma (t) . \label{eq83}
\end{equation}
The coefficient $\sigma (t)$ contains $\gamma(t)$ due to   $G(x)$ in Eq. (\ref{eq1P}). In order to obtain   a quantity which depends only on the time-dependent diffusion coefficient  we should take  
$\gamma (t)=0$.
 Then, the first generalized  moment describes only the deterministic part  of the system, whereas the generalized variance  is related to the fluctuations of the system. Fig. \ref{fig4} shows  the simulated and analytical results for the relation  (\ref{eq80bb}) and the generalized variance described by Eq.  (\ref{eq83}); they are in good agreement. The plateaus are due to $G(x)$ in Eq.  (\ref{eq1P}).

It is interesting to note that the ordinary $n$-moment can  be written in terms of the first  generalized moment and generalized variance as follows:
\begin{equation}
 \left< x^n\right> =e^{n\left< G(x)\right>+\frac{n^2\left< \left( G(x)-\left< G(x)\right>\right)^2\right>}{2}} . \label{eq84}
\end{equation}

Now we consider two  applications of the Langevin equation (\ref{eq1P}) with non-constant coefficients. In the first one we take  the application of the generalized variance  for a reformulation of the Verhulst logistic diffusion model described in Ref. \cite{roman} by using the time-space-dependent drift coefficient as follows:
\begin{equation}
\frac{dx}{dt}= \frac{bc}{b+e^{ct}}x+\bar{\sigma} x L(t)  , \ \  0< x< \infty . \label{eq91}
\end{equation}%
Eq. (\ref{eq91}) is a Langevin equation in the Ito prescription which has been used to describe a culture of microorganisms \cite{roman}. However, the solution (\ref{eq15P}) has been obtained in the Stratonovich prescription. The corresponding Langevin equation  (\ref{eq1P})  to the Ito prescription  is given by
\begin{equation}
\frac{dx}{dt}=b\left( \frac{c}{b+e^{ct}}-\frac{\bar{\sigma}^2}{b}\right) x+\bar{\sigma} x L(t)  , \label{eq92}
\end{equation}%
which has an additional term in the drift term.

In this case, the drift and diffusion coefficients of Eq.  (\ref{eq1P}) are given by  $b(t)=\bar{\sigma} $, $\beta (t)=bc/\left( b+e^{ct}\right) -\bar{\sigma}^2$, $\gamma (t)=0$ and $D(x)=x$.   The normalization of the PDF implies that $x$ should also be restricted to $0< x< \infty $ and we obtain
\begin{equation}
\rho \left( x,t\right) =
\frac{e^{  -\frac{\left( \ln \left( \frac{x}{x_0}\right)-\left( c-\bar{\sigma}^2\right) t -\ln \left( \frac{1+b}{b+e^{c t}}\right) \right)^2}{4 \bar{\sigma}^2 t} }} {\sqrt{4\pi \bar{\sigma}^2t}x}. \label{eq93}
\end{equation}
The ordinary $n$-moment can be calculated from Eq. (\ref{eq80}) and it is given by
\begin{equation}
 \left< x^n\right> =\left[ \frac{x_0(1+b)}{1+be^{-ct}}\right]^ne^{-n(1-n)\bar{\sigma}^2t} . \label{eq94}
\end{equation}
For the first moment we have
\begin{equation}
 \left< x\right> = \frac{x_0(1+b)}{1+be^{-ct}} . \label{eq95}
\end{equation}
Eq.  (\ref{eq95}) has the same result reported in Ref. \cite{roman}. We can see that the first moment  (\ref{eq95}) does not depend on the  parameter $\bar{\sigma}$ due to the fact that the drift coefficient has the term (-$\bar{\sigma}^2 x$) which cancels the spurious drift term generated from the Stratonovich prescription. A quantity related to the   diffusion coefficient can be described by Eq. (\ref{eq80c}),
\begin{equation}
 \frac{\left< x^2\right> }{\left< x\right>^2}=e^{2\bar{\sigma}^2 t} . \label{eq96}
\end{equation}
We can  quantify the  fluctuations of the noise by using the generalized variance given by
\begin{equation}
\left< \left(\ln (x)-\left< \ln (x)\right>\right)^2\right> =2\bar{\sigma}^2t  . \label{eq97}
\end{equation}%
It can be seen that the generalized variance has a more simple form to quantify the noise than the expression (\ref{eq96}).

The second application of the system (\ref{eq1P})  is on the impact of chloride ion diffusion in concrete \cite{liu,yan}. The studies indicated that chloride ion diffusion in concrete is time-dependent \cite{yan}. The model used in Ref. \cite{yan} is, basically, a diffusion equation with time-dependent diffusion coefficient described by
\begin{equation}
\frac{\partial \rho(x,t)}{\partial t}=Db_{\eta}t^{\eta} \frac{\partial^2\rho (x,t) }{\partial x^2}  ,  \label{eq97}
\end{equation}%
where $\rho(x,t)$ describes the concentration of the chloride ions. The model  (\ref{eq97}) can be described in terms of the Langevin equation  (\ref{eq1P}) with $\beta =\gamma =0$, $D(x)=1$ and $b(t)=Db_{\eta}t^{\eta}$.

\subsection{Generalized Einstein relation}

A connection between the fluctuations of an
ensemble of particles and  their mobility under an applied small load force is described by the second Einstein relation which connects the first moment in the presence of a constant force to the second moment without any external force. In this case, the PDF is described by a Gaussian distribution. The Einstein relation has been  generalized for the cases of non-Gaussian distributions described by a wide class of non-linear Langevin equations \cite{kwokERc}. Now we generalize the Einstein relation to the system (\ref{eq1P}). We first consider  the first generalized  moment given by Eq.  (\ref{eqcase63}).
Then we connect  the first generalized moment to the generalized variance  (\ref{eq83}) as follows:
\begin{equation}
\left< \left( G(x)-G_0e^{\int_{0}^{t} \text{d}u \gamma (u)}\right)\right> =\left< \left( G(x)-\left< G(x)\right>\right)^2\right>  \label{eqER1b}
\end{equation}%
 by taking
\begin{equation}
\int_{0}^{t} \text{d}z \beta (z) e^{\int_{z}^{t} \text{d}u \gamma (u)}=2\int_{0}^{t} \text{d}z b^2(z) e^{2\int_{z}^{t} \text{d}u \gamma(u)} . \label{eqER2}
\end{equation}%
As example, we take $\beta (t)=b^2( 1+e^{2\gamma t})$, $\gamma (t)=\gamma$ and $b(t)=b$; they satisfy the relations (\ref{eqER1b}) and (\ref{eqER2}).

However, for the generalized variance (\ref{eq83})  to describe only the fluctuations of the noise we should take $\gamma (t)=0$, then we have
\begin{equation}
\int_{0}^{t} \text{d}z \beta (z) e^{\int_{z}^{t} \text{d}u \gamma (u)}=2\int_{0}^{t} \text{d}z b^2(z)  , \label{eqER2b}
\end{equation}%
and the  generalized second Einstein relation for the system (\ref{eq1P}) is described by
\begin{equation}
\left< \left( G(x)-G_0e^{\int_{0}^{t} \text{d}u \gamma (u)}\right)\right>_{\gamma} =\left< \left( G(x)-\left< G(x)\right>\right)^2\right>_{\gamma =0} . \label{eqER3}
\end{equation}%
As example, we take $\beta (t)=2b^2(1-\gamma t)$, $\gamma (t)=\gamma$ and $b(t)=b$; they satisfy the relations (\ref{eqER2b}) and (\ref{eqER3}).

\section{ Conclusion}

A wide class of  nonlinear Langevin equations with the drift and diffusion coefficients separable in time and space has been considered, and they can be applied to various systems. We have obtained  solutions for the generalized correlation function (\ref{eqTAM5b}) and GETAMSD (\ref{eqTAM5c}). As is well-known, the Langevin equation with space-time-dependent coefficients may describe complicated behaviors, then a convenient choice of the  observable may be useful to simplify the results of $n$-moment and also lead to some new relations and properties. We have shown that the observable $G(x)$ is useful to describe the system  (\ref{eq1P}) with  space-time-dependent drift and diffusion coefficients in the Stratonovich prescription:  It can be used to obtain a generalized Einstein relation and  attain ergodicity, and it is also useful to calculate the mean generalized squared displacement of the fluctuations of the system by using the generalized variance.  A new relation which connects the ordinary $n$-moment with the first two generalized moments has also been obtained (\ref{eq84}).

Possibly, as an extension of this work, we may also use the observable $G(x)$ or other generalized observables  to calculate generalized moments  of  more complicated systems described by 
\begin{equation}
\frac{dx}{dt}= h(x,t)+g(x,t) L(t)  ,  \label{eqCL1}
\end{equation}%
where $h(x,t)$ and $g(x,t)$ are some class of generic functions which depend on the space-time coordinates. See Ref. \cite{kwokAsym} for a system different from Eq. (\ref{eq1P}) in which  the observable $G(x)$ has been employed for calculating the generalized $n$-moment.



\end{document}